%% file: main.tex
\pdfoutput=1
\documentclass[conference]{IEEEtran}

\usepackage[utf8]{inputenc}
\usepackage[english]{babel}

\usepackage{csquotes}

\usepackage{amsmath, amssymb,wasysym}

\usepackage{graphicx,color}
\usepackage[dvipsnames]{xcolor}
\usepackage{float}
\usepackage{subfloat,subcaption}

\usepackage{tikz}
\usetikzlibrary{automata,arrows,shadows}

\usepackage{tabularx}
\usepackage{makecell}
\usepackage{multirow}
\usepackage{colortbl}

\usepackage{listings}
\lstdefinestyle{vhdl_style}
{
    language=VHDL,
    float=!htb,
    basicstyle=\ttfamily\footnotesize,
    identifierstyle=\bfseries\color{black},
    keywordstyle=\bfseries\color{OrangeRed!80},
    stringstyle=\bfseries\color{yellow},
    commentstyle=\bfseries\color{gray},
    columns=flexible,
    frame=single,
    showspaces=false,
    showstringspaces=false,
    numberstyle=\tiny,
    stepnumber=1,
    breaklines=true,
    xrightmargin=-\fboxsep,
    backgroundcolor=\color{white},
    captionpos=t,
    mathescape,
    escapechar=\%
}

\usepackage{algorithm}
\usepackage[noend]{algpseudocode}
\algrenewcommand\algorithmicrequire{\textbf{\ \ Input:}}
\algrenewcommand\algorithmicensure{\textbf{Output:}}

\usepackage{url}
\usepackage[bookmarks=false]{hyperref}

\usepackage{microtype}
\usepackage{xspace}

\usepackage{comment}

\usepackage[nolist]{acronym}
\input{acronyms.tex}

\newcommand{\CC}[1][]{$\text{C\hspace{-.25ex}}^{_{_{_{++}}}}\ifthenelse{\equal{#1}{}}{}{\text{\hspace{-.625ex}#1}}$}

\newcommand{\circled}[2][]{
  \tikz[baseline=(char.base)]{
    \node[shape=circle,draw,inner sep=1pt,fill=black]
    (char) {\phantom{\ifblank{#1}{#2}{#1}}};
    \node[text=white] at (char.center) {\makebox[0pt][c]{\bf #2}};}}
\robustify{\circled}

\newcommand{\etal}[0]{\textit{et~al.}\xspace}
\newcommand{\romannumber}[1]{\uppercase\expandafter{\romannumeral#1}}

\newcommand{\Algorithm}[1]{\textcolor{blue}{Alg.~\ref{#1}}}

\newcommand{\Figure}[1]{\textcolor{blue}{Fig.~\ref{#1}}}

\newcommand{\Section}[1]{\textcolor{blue}{Sect.~\ref{#1}}}

\begin{document}

\title{
         A Look at the Dark Side of Hardware Reverse Engineering -- A Case Study
}

\author{
        \IEEEauthorblockN{
                Sebastian Wallat\IEEEauthorrefmark{1},
                Marc Fyrbiak\IEEEauthorrefmark{2},
                Moritz Schlögel\IEEEauthorrefmark{2},
                Christof Paar\IEEEauthorrefmark{1}\IEEEauthorrefmark{2}, \IEEEmembership{Fellow,~IEEE}\\
        }
        \IEEEauthorblockA{
                \IEEEauthorrefmark{1}University of Massachusetts Amherst, USA\\
                \IEEEauthorrefmark{2}Horst G\"ortz Institute for IT Security, Ruhr University Bochum, Germany\\
                \textit{swallat}@umass.edu,
                \textit{prename.surname}@rub.de,\\
        }
}

\maketitle

\input{section/abstract.tex}


\input{section/introduction}
\input{section/technical_introduction}

\input{section/case_study_wm}
\input{section/case_study_block_ciphers}
\input{section/case_study_a51}
\input{section/conclusion}

\section*{Acknowledgment}
This material is based upon work partially supported by NSF award CNS-1563829 and ERC grant 695022.

\bibliographystyle{IEEEtran}
{\small
        \bibliography{bibliography}
}

\end{document}

%% file: acronyms.tex
\begin{acronym}

\acro{3DES}{Triple-DES}

\acro{AES}{Advanced Encryption Standard}
\acro{ALU}{Arithmetic Logic Unit}
\acro{API}{Application Programming Interface}
\acro{ARX}{Addition Rotation XOR}
\acro{ASIC}{Application Specific Integrated Circuit}
\acro{ASIP}{Application Specific Instruction-Set Processor}
\acro{AS}{Active Serial}

\acro{BDD}{Binary Decision Diagram}
\acro{BGL}{Boost Graph Library}
\acro{BNF}{Backus-Naur Form}
\acro{BRAM}{Block-Ram}

\acro{CBC}{Cipher Block Chaining}
\acro{CFB}{Cipher Feedback Mode}
\acro{CFG}{Control Flow Graph}
\acro{CLB}{Configurable Logic Block}
\acro{CLI}{Command Line Interface}
\acro{COFF}{Common Object File Format}
\acro{CPA}{Correlation Power Analysis}
\acro{CPU}{Central Processing Unit}
\acro{CRC}{Cyclic Redundancy Check}
\acro{CTR}{Counter}

\acro{DC}{Direct Current}
\acro{DES}{Data Encryption Standard}
\acro{DFA}{Differential Frequency Analysis}
\acro{DFS}{Depth-First Search}
\acro{DFT}{Discrete Fourier Transform}
\acro{DLL}{Dynamic Link Library}
\acro{DMA}{Direct Memory Access}
\acro{DNF}{Disjunctive Normal Form}
\acro{DOD}[DoD]{Department of Defense}
\acro{DPA}{Differential Power Analysis}
\acro{DSO}{Digital Storage Oscilloscope}
\acro{DSP}{Digital Signal Processing}
\acro{DUT}{Device Under Test}
\acro{DUA}{Device Under Analysis}

\acro{ECB}{Electronic Code Book}
\acro{ECC}{Elliptic Curve Cryptography}
\acro{EEPROM}{Electrically Erasable Programmable Read-only Memory}
\acro{EMA}{Electromagnetic Emanation}
\acro{EM}{electro-magnetic}

\acro{FFT}{Fast Fourier Transformation}
\acro{FF}{Flip Flop}
\acro{FI}{Fault Injection}
\acro{FIR}{Finite Impulse Response}
\acro{FPGA}{Field Programmable Gate Array}
\acro{FSM}{Finite State Machine}

\acro{GUI}{Graphical User Interface}

\acro{HDL}{Hardware Description Language}
\acro{HD}{Hamming Distance}
\acro{HF}{High Frequency}
\acro{HSM}{Hardware Security Module}
\acro{HW}{Hamming Weight}

\acro{IC}{Integrated Circuit}
\acro{IO}[I/O]{Input/Output}
\acro{IOB}{Input Output Block}
\acro{IOT}[IoT]{Internet of Things}
\acro{IP}{Intellectual Property}
\acro{ISA}{Instruction Set Architecture}
\acro{IV}{Initialization Vector}

\acro{JTAG}{Joint Test Action Group}

\acro{KAT}{Known Answer Test}

\acro{LFSR}{Linear Feedback Shift Register}
\acro{LSB}{Least Significant Bit}
\acro{LUT}{Look-up table}

\acro{MAC}{Message Authentication Code}
\acro{MIPS}{Microprocessor without Interlocked Pipeline Stages}
\acro{MMIO}{Memory Mapped \acl{IO}}
\acro{MSB}{Most Significant Bit}

\acro{NASA}{National Aeronautics and Space Administration}
\acro{NSA}{National Security Agency}
\acro{NVM}{Non-Volatile Memory}

\acro{OFB}{Output Feedback Mode}
\acro{OISC}{One Instruction Set Computer}
\acro{ORAM}{Oblivious Random Access Memory}
\acro{OS}{Operating System}

\acro{PAR}{Place-and-Route}
\acro{PCB}{Printed Circuit Board}
\acro{PC}{Personal Computer}
\acro{PS}{Passive Serial}
\acro{PUF}{Physical Unclonable Function}

\acro{RAM}{Random Access Memory}
\acro{RISC}{Reduced Instruction Set Computer}
\acro{RNG}{Random Number Generator}
\acro{ROM}{Read-Only Memory}
\acro{ROP}{Return-oriented Programming}
\acro{RTL}{Register Transfer Level}

\acro{SBOX}[S-Box]{Substitution-Box}
\acrodefplural{SBOX}[S-Boxes]{Substitution-Boxes}
\acro{SCA}{Side-Channel Analysis}
\acro{SHA}{Secure Hash Algorithm}
\acro{SNR}{Signal-to-Noise Ratio}
\acro{SOC}[SoC]{System-on-Chip}
\acro{SPA}{Simple Power Analysis}
\acro{SPN}{Substitution Permutation Network}
\acro{SPI}{Serial Peripheral Interface Bus}
\acro{SRAM}{Static Random Access Memory}

\acro{UART}{Universal Asynchronous Receiver Transmitter}
\acro{UHF}{Ultra-High Frequency}
\acro{USB}{Universal Serial Bus}

\acro{VHDL}{Very High Speed Integrated Circuit Hardware Description Language}

\acro{WISC}{Writeable Instruction Set Computer}

\acro{XDL}{Xilinx Description Language}
\acro{XTS}{XEX-based Tweaked-codebook with ciphertext Stealing}

\end{acronym}

%% file: section/abstract.tex

\begin{abstract}

A massive threat to the modern and complex \acs{IC} production chain is the use of untrusted off-shore foundries which are able to infringe valuable hardware design \acs{IP} or to inject hardware Trojans causing severe loss of safety and security. Similarly, market dominating \acs{SRAM}-based \acsp{FPGA} are vulnerable to both attacks since the crucial gate-level netlist can be retrieved even in field for the majority of deployed device series. In order to perform \acs{IP} infringement or Trojan injection, reverse engineering (parts of) the hardware design is necessary to understand its internal workings. Even though \acs{IP} protection and obfuscation techniques exist to hinder both attacks, the security of most techniques is doubtful since realistic capabilities of reverse engineering are often neglected.

The contribution of our work is twofold:
first, we carefully review an \acs{IP} watermarking scheme tailored to \acsp{FPGA} and improve its security by using opaque predicates. In addition, we show novel reverse engineering strategies on proposed opaque predicate implementations that again enables to automatically detect and alter watermarks.
Second, we demonstrate automatic injection of hardware Trojans specifically tailored for third-party cryptographic \acs{IP} gate-level netlists. More precisely, we extend our understanding of adversary's capabilities by presenting how block and stream cipher implementations can be surreptitiously weakened.

\end{abstract}

%% file: section/introduction.tex

\section{Introduction}
\label{Trojans:section:introduction}

To assure security and safety of any modern application, it is of crucial importance that the underlying cryptographic primitive is not compromised. Given that most crypto algorithms in use such as the Suite B ciphers~\cite{nist:suite:b} are robust against traditional attacks, i.e. brute-force and cryptanalysis, adversaries are often forced to exploit implementation attacks. The most prominent examples of implementation attacks are \ac{SCA} and \ac{FI}. Both attack families have been investigated in great detail in the \ac{ASIC} and \ac{FPGA} context over the last two decades, in the scientific community as well as in industry, cf.~\cite{drimer-thesis,cosade:2015:wild,Sasdrich2014,7426148,5513117}. Even though \ac{SCA} and \ac{FI} countermeasures are certainly not solved problems, there is a sound understanding of attacks and countermeasures.

Another prominent class of attacks attempt to weaken the security by manipulating the underlying hardware, often referred to as Trojans. In spite of extensive research~\cite{karri_hstrust_2012,SurveyTrojan}, there are still open questions regarding the practicability of many proposed Trojans since either adversarial access to the source code is assumed~\cite{5406669} or crucial reverse engineering steps are neglected~\cite{King:2008:DIM:1387709.1387714}. However, in real-world scenarios the adversary (e.g., a malicious foundry) faces daunting tasks of: (1) reverse engineering high-level information from a gate-level netlist, (2) overcoming possible \ac{IP} protection mechanisms, (3) identifying the security-critical modules, (4) followed by the actual Trojan insertion in the target design. Understanding the feasibility and complexity of these steps is crucial for a sound estimation of the threat posed by hardware manipulations and more importantly, aid with developing sound countermeasures against Trojans.

\textbf{Goals and Contributions.}
In this paper, we focus on destructive aspects of reverse engineering. Our goal is to demonstrate the practicability of \ac{IP} infringement and hardware Trojan injection for third-party gate-level netlists of market-dominating \ac{SRAM}-based \acp{FPGA}. To this end, we first review the security of a constraint-based \ac{IP} watermarking scheme specifically tailored for \acp{FPGA} and subsequently we show general improvements to increase the security against reverse engineering.
We then demonstrate the automation of hardware Trojan injection in third-party gate-level netlists of cryptographic designs. To highlight devastating consequences of hardware Trojans for both block ciphers and stream ciphers, we selected the standardized and widely-used Present block cipher and A5/1 stream cipher.
Our main contributions are:
\begin{itemize}
    \item {\bf \ac{IP} Infringement.}
    In our first case study (\Section{hwre:section:case_wm}), we carefully analyze the security of constraint-based watermarking tailored to \acp{FPGA} and show how to automatically identify and tamper watermarks. Additionally, we present novel improvements to mitigate reverse engineering by use of opaque predicates. Moreover, we demonstrate flaws in proposed hardware opaque predicate implementations.

    \item {\bf Hardware Trojan Injection.}
    In our second case study (\Section{trojans:section:block_ciphers} and \Section{trojans:section:stream_ciphers}), we demonstrate how benign designs of Present and A5/1 can be surreptitiously weakened. In particular, we provide novel aspects on how these cryptographic algorithms can be automatically reverse engineered and custom-tailored key leakage mechanism that allow a man-in-the-middle eavesdropper to decrypt any communication.
\end{itemize}

%% file: section/technical_introduction.tex

\section{Technical Background and Related Work}
\label{trojans:section:tech_intro_related_work}

Our work builds on previous research in \ac{IP} protection, reverse engineering, and hardware Trojans. Below, we present a concise technical background and related work on each area.

\subsection{IP Protection}
\label{trojans:sub_section:ip_infr_water}

The design of digital systems involves valuable effort in terms of manpower and money. To reduce both production time and costs, the reuse of \ac{IP} design components is a common practice. Unfortunately, the protection of the \ac{IP} owner's rights becomes a major problem~\cite{ieee:2014:guin}. In recent years, various solutions have been proposed to protect valuable \ac{IP}~\cite{7082768}. Generally, there exists numerous defenses (e.g., watermarking or fingerprinting) targeting different attacks (e.g., overbuilding or cloning).

\subsection{Hardware Reverse Engineering}
\label{trojans:sub_section:gate_level_rev_eng}

Hardware reverse engineering can be generally divided in \ac{FPGA} and \ac{ASIC}-oriented techniques. In the reverse engineering context of \acp{FPGA} existing works can be mainly divided into bitstream reverse engineering to recover the netlist, and netlist reverse engineering to recover high-level \ac{RTL} information.
Obtaining of the bitstream from a deployed \ac{SRAM}-based \ac{FPGA} requires to extracting it from the storing non-volatile memory. While several works developed automated file format reverse engineering techniques to recover the (partial) bitstream, c.f.~\cite{book:security_trends_fpga:chapter2}, there also exist works on directly selecting the bitstream as an target for manipulations, c.f.~\cite{Swierczynski:CAIDI:2015,bifi}. Although prominent \ac{FPGA}s provide bitstream encryption to achieve confidentiality, it could be shown that in many cases this encryption can be circumvented~\cite{moradi:2011:ccs,moradi:2013:fpga}.
In the case of \ac{ASIC} netlist reverse engineering several different areas of research can be divided. Where one area focuses on recovering high-level extraction~\cite{dt:1999:chisholm,shi:2010:iscas,li:2013:host,gascon:2014:fmcad}, other works elaborate on best practices for a human analyst~\cite{hansen:1999:dt}.

\subsection{Hardware Trojans}
\label{trojans:sub_section:hw_trojans}

After the publication of the initial \ac{DOD} report~\cite{DSB-Report} the research community started to investigate both offensive and defensive aspects of malicious hardware manipulations~\cite{karri_hstrust_2012,SurveyTrojan}.

Defense-driven research focuses on the detection of hardware Trojans using several characteristic properties~\cite{fanci,veritrust}. Albeit, Zhang~\etal~\cite{detrust} were able to present a solution to evade detection algorithms targeting the netlist.
In contrast to the defense-driven side much less work has been performed on the attack-driven side. The majority of proposed Trojan designs are inserted during the design phase while accessing high-level information~\cite{King:2008:DIM:1387709.1387714}. Furthermore there are several works target novel Trojan design methodologies such as dopant-level Trojans~\cite{becker_ches_2013}, analog malicious hardware~\cite{oakland:2016:yang}, or parametric Trojans~\cite{ghandali:2016:ches}.

\subsection{Attacker Model}
\label{trojans:sub_section:attacker_model}

We suppose that the adversary has access to a flattened gate-level netlist without any high-level information such as names, module hierarchies, and synthesis options. The goal of the adversary is to perform an illegitimate application such as hardware Trojan injection or \ac{IP} infringement. In consequence the attacker has to at least partially reverse-engineer the design.

Note that our adversary model is realistic in multiple scenarios, i.e. \ac{SRAM}-based \acp{FPGA} (since the netlist can be recovered from a bitstream~\cite{DingWZZ13}) and malicious foundries.

%% file: section/case_study_wm.tex

\section{Case Study: Watermarking}
\label{hwre:section:case_wm}

In this case study, we present how \ac{IP} infringement can be performed for watermark protected designs. To demonstrate the efficacy of gate-level netlist reverse engineering, we analyze a scheme which aims to protect valuable \ac{FPGA} \ac{IP} cores at netlist level.

\subsection{LUT-based Watermarking}
As described in \Section{trojans:section:tech_intro_related_work}, various works have addressed \ac{IP} protection by use of watermarks. In particular, constraint-based watermarking~\cite{tcadics:2001:kahng} is suited for \acp{FPGA} since the additional satisfiability constraints are suited for the \ac{LUT}-based \ac{FPGA} structures~\cite{ftp:2008:schmid}.

\par
\begin{figure}[!htb]
  \centering
  \includegraphics[scale=0.75]{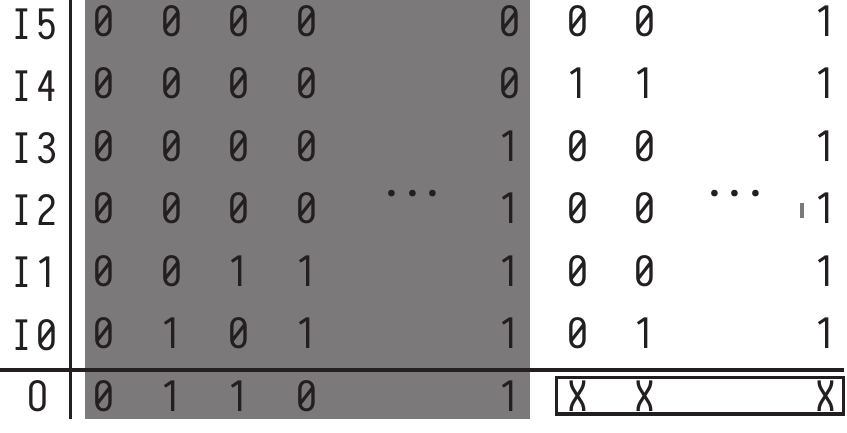}
  \caption{Example of a LUT-6. I0 to I5 are input pins, and O is the output pin. I4 and I5 are connected to \textsf{GND}. Bits marked with \textsf{X} can be used for watermarks.}
  \label{figure:watermarking}
\end{figure}

{\bf Watermarking Scheme~\cite{ftp:2008:schmid}.}
The high-level idea of the scheme by Schmid~\etal\ is to exploit not addressable \ac{LUT} memory space,  see~\Figure{figure:watermarking}. Since input pin I1 is connected to \textsf{GND}, there are 48 Bits that can be arbitrarily changed without altering any functionality of the design. These \ac{LUT} memory bits are used to embed the watermark.

\par{\bf Reverse Engineering.}
In order to identify the \ac{LUT} memory bits which implement the watermark, we analyze the Boolean function of each \ac{LUT}. More precisely, we inspect each clause of a \ac{LUT}'s \ac{DNF}. If there is any clause where an \textsf{GND} or \textsf{VCC} input signal is required to be logical \textsf{1} or \textsf{0}, respectively, we successfully identified a watermark bit. We practically verified that we are able to automatically disclose the watermark of our target design (an \ac{AES} \ac{IP} core synthesized for an \textsf{xc6slx16} with Xilinx ISE 14.7). Furthermore, we are able to automatically remove and alter the watermark in the gate-level netlist.

Note that Schmid~\etal~\cite{ftp:2008:schmid} proposed to utilize  \acp{LUT} configured as shift register \acp{LUT} to prevent  optimization, but since the \textit{shift-enable} input pin is  assigned to \textsf{GND} (to ensure the designs functionality), we treat them as general \acp{LUT} in our reverse engineering algorithm.

\subsection{Opaque Predicates}
Even though Schmid~\etal~\cite{ftp:2008:schmid} noted that the security can be increased by use of bogus constant generating signals (instead of the plain connection to \textsf{GND}), it appears challenging how such signals can be implemented with consideration of a reasonable reverse engineer.

To this end, we propose to leverage \textit{opaque predicates} which have been mainly used in the context of software watermarking and obfuscation~\cite{ecr:2006:ginger}. An opaque predicate is an expression that either evaluates to \textit{true} or \textit{false} irregardless of the input and thus this function implements a constant generating signal. Hence, instead of \textsf{GND} the \textit{unused} \ac{LUT}'s input pins (e.g., I5 and I6 in \Figure{figure:watermarking}) are connected to the output of an opaque predicate which mitigates our reverse engineering approach presented in the previous section. Despite numerous works that address opaque predicates for software, there is to the best of our knowledge only one work by Sergeichik~\etal~\cite{sergeichik:2014:jicms} applying them to the hardware reverse engineering context.

\par{\bf Implementation~\cite{sergeichik:2014:jicms}.}
To implement opaque predicates, Sergeichik~\etal\ suggest to exploit \acp{LFSR} as constant signal generators. Their general idea is to connect all registers of the \ac{LFSR} to an \textsf{OR} or \textsf{NOR} gate to generate a constant high or low signal,  respectively. Note that the standard \ac{LFSR} with \textsf{XOR} feedback never enter the zero state (where all registers hold a logic \textsf{0}) thus there is at least one register which holds a logic~\textsf{1}.

\par{\bf Reverse Engineering.}
Even though the reverse engineering strategy from the previous section is generally prevented by opaque predicates, the proposed opaque predicate instantiation by \acp{LFSR} is not sufficient to mitigate reverse engineering and thus \ac{IP} infringement is again possible. The high-level idea of our automated reverse engineering strategy is to identify \acp{LFSR} and subsequently whether they are used to implement constant generators. For the detection of \acp{LFSR}, we exploit their typical characteristic: a chain of \acp{FF} elements to store and shift the current state. Note that we skip any pass-through \acp{LUT} or buffers in each step.
To find the \ac{FF} chains, we search for the \textit{initial} \ac{FF} (which stores the next state bit) by checking whether its preceding gate is a \ac{FF}. For any candidate, we execute a modified \ac{DFS}, and we search for circles considering taps defined by the underlying feedback polynomial of the \ac{LFSR}. Since we can identify the position of the initial \ac{FF} and the taps, we are able to algorithmically identify the feedback polynomial as well.

After the \acp{LFSR} are automatically reverse engineered, we topologically analyze each \ac{LFSR} for the constant generator part, i.e. an \textsf{OR} or \textsf{NOR} gate that is connected to all registers of the \ac{LFSR}. Note that the \textsf{OR} and \textsf{NOR} gates might be implemented across multiple \acp{LUT} depending on the type and size of the \ac{LFSR}. To identify the final constant generation gate, we manually inspected our target design (an \ac{AES} \ac{IP} core synthesized for an \textsf{xc6slx16} with Xilinx ISE 14.7) which typically requires only several minutes but can be easily automated.
In case other types of \acp{LFSR} such as \textsf{XNOR}-based \acp{LFSR}~\cite{sergeichik:2014:jicms}, the search for the final constant generation gate has to be adapted.

In summary, \ac{LUT}-based watermarking is a promising approach to protect valuable \ac{FPGA} \ac{IP} and particularly in combination with advanced hardware obfuscation techniques it might provide an adequate security level to hamper reverse engineering.

%% file: section/case_study_block_ciphers.tex
\section{Case Study: Block Ciphers}
\label{trojans:section:block_ciphers}

This case study presents an approach to detect and weaken block ciphers using \acp{SBOX}. Here, we extend the technique described in Swierczynski~\etal~\cite{Swierczynski:CAIDI:2015} by weakening block ciphers normally found in space and power constraint \ac{IOT} devices on the netlist level. By an example we describe the reverse engineering and Trojan injection process for an \ac{FPGA} based design for an Xilinx Spartan-6 device. Using PRESENT~\cite{Bogdanov:2007:PUB:1421964.1422007} as the targeted design we demonstrate an fully algorithmic solution to detect (\Section{trojans:section:block_ciphers:detect}) and weaken (\Section{trojans:section:block_ciphers:trojan}) the \acp{SBOX} of the implementation, even if the \ac{SBOX} logic is merged with the control logic of a design.

\textbf{Block Ciphers.} In modern systems block ciphers like \ac{AES} are the standard for symmetrically encrypted communication. In contrast to asymmetric ciphers, many prominent symmetric schemes use a \ac{SPN} structure which allows a higher data throughput. In combination with asymmetric schemes, which enable the on-demand key generation over an untrusted channel, they allow the encryption of plaintext in blocks of several bytes. In contrast to stream ciphers, where each bit is encrypted independently, block ciphers use a predefined block size to specify the amount of data that is encrypted at once. The \ac{SPN} structure consists here of alternative substitutions and permutations. The characteristic non-linear nature of each \ac{SBOX} leads to a specific representation in the synthesized \ac{FPGA} netlist.

\textbf{PRESENT cipher.} Although \ac{AES} is the de-facto standard for block-ciphers, due to it size it has some drawbacks in constrained scenarios, as smart-cards, medical, or battery powered \ac{IOT} devices. One prominent cipher to use in this context is the PRESENT cipher. In contrast to \ac{AES}, which uses a 8-bit to 8-bit \ac{SBOX}, it uses a 4-bit to 4-bit \ac{SBOX} while increasing the number of rounds. For LUT-6 architectures, \acp{LUT} with six inputs and (usually) one output, like Spartan-6 the \ac{SBOX} logic can be potentially merged by the synthesizer. Here two input pins are not used by the \ac{SBOX} logic and can therefore be freely used by the synthesizer to reduce the space consumption of a design.

In the following we discuss our approach based on Swierczynski~\etal~\cite{Swierczynski:CAIDI:2015} of detecting \acp{SBOX} in \ac{FPGA} bitstreams, and extend it to detect merged \acp{SBOX} on the netlist level. Therefore we facilitate an existing PRESENT implementation~\cite{present-opencore} and utilize Xilinx ISE 14.7 to produce a gate-level, \ac{HDL}-based netlist of the implementation. Such a modification can be of interest to extract the secret key from a \ac{DUT}, like an \ac{USB} flash drive, discussed in~\cite{jcen:2016:swierczynski}.

\begin{algorithm}[!htb]
    \caption{PRESENT \ac{SBOX} pattern generation for a LUT-6 \ac{FPGA} architecture}
    \begin{algorithmic}[1]
        \item[]{\textbf{Input}:~Present \ac{SBOX} \textit{sbox}}
        \item[]{\textbf{Input}:~Number of input bits used for the \ac{SBOX} \textit{l}}
        \item[]{\textbf{Output}:~Vector of all configuration for PRESENT~\acp{LUT} \textit{v}}
        \item[]{\hrulefill}
        \item[]{Map of all permutation to the generated configuration vector \textit{m}}
        \item[]{\ac{SBOX} value at position n \textit{sbox[n]}}
        \item[]{\hrulefill}
        \For{\textit{$i := 0; i < 2^l; ++i$}}
            \For{\textit{p} in Bitpermutations of \textit{i}}
                \State{\textit{$m[p].push\_back(sbox[p])$}}
            \EndFor
        \EndFor
        \State{\textit{return $m.values()$}}
    \end{algorithmic}
    \label{alg:present_pattern_generation}
\end{algorithm}

\begin{algorithm}[!htb]
    \caption{\ac{LUT} Sub-Configuration Generator for \ac{FPGA} based designs}
    \begin{algorithmic}[1]
        \item[]{\textbf{Input}:~Configuration to decompose \textit{orig\_conf}}
        \item[]{\textbf{Input}:~Vector of input pins to sequentially remove from configuration: \textit{p}}
        \item[]{\textbf{Output}:~Vector of all generated Sub-Configurations \textit{s}}
        \item[]{\textbf{Output}:~Vector of pins in order of removal \textit{v}}
        \item[]{\hrulefill}
        \item[]{Vector of Configuration: \textit{r}}
        \item[]{Number of Input Pins for Configuration \textit{c}: \textit{$c.input\_pin\_size()$}}
        \item[]{Access \textit{i'th} Bit of Number \textit{n} (MSB at position: 0): $n[i]$}
        \item[]{\hrulefill}
        \State{\textit{$r.push\_back(orig\_conf)$}}
        \While{!p.empty()}
            \State{\textit{$pin := p.take\_first()$}}
            \State{\textit{$v.push\_back(pin)$}}
            \State{Intermediate Vector of Configurations \textit{$r\_tmp := r$}}
            \For{\textit{c} in \textit{r\_tmp}}
                \State{Configuration Vector $\mu_1$, $\mu_2$}
                \State{$l := c.input\_pin\_size()$}
                \For{\textit{$n = 0; n < 2^{l}; ++n$}}
                    \If{\textit{n[p] == 0}}
                        \State{\textit{$\mu_0.push\_back(c[n])$}}
                    \Else
                        \State{\textit{$\mu_1.push\_back(c[n])$}}
                    \EndIf
                \EndFor
                \State{\textit{$r\_tmp.push\_back(\mu_0)$}}
                \State{\textit{$r\_tmp.push\_back(\mu_1)$}}
            \EndFor
            \State{\textit{$r := r\_tmp$}}
        \EndWhile
        \State{\textit{return $r, v$}}
    \end{algorithmic}
    \label{alg:lut_shrink}
\end{algorithm}

\subsection{\ac{SBOX} detection}
\label{trojans:section:block_ciphers:detect}

In contrast to \ac{AES}, where one LUT-6 is not able to represent the calculation of one \ac{SBOX} output bit, PRESENT using a 4-to-4 \ac{SBOX} allows the implementation in one \ac{LUT} and also allows two bits to be freely used by the synthesizer. \Algorithm{alg:present_pattern_generation} provides an approach to generate the LUT-4 based output pattern, \acp{LUT} with four inputs and one output, for the PRESENT cipher. In order to search for these pattern in the LUT-6 based netlist we iterate over all LUT elements and need to extract the sub-configurations by sequentially removing one input bit from the \ac{LUT} equation. Here we iterate over all combinations of two input pin tuples and generate the sub-pattern, c.f. \Algorithm{alg:lut_shrink}. In our case we were able to detect 68 \ac{SBOX} implementation, 64 for each bit of a state and four used by the key-schedule.

\begin{algorithm}[!htb]
    \caption{\ac{LUT} Sub-configuration Merger for \ac{FPGA} based designs}
    \begin{algorithmic}[1]
        \item[]{\textbf{Input}:~Vector of Sub-Configurations \textit{s}}
        \item[]{\textbf{Input}:~Vector of pins in order of removal \textit{v}}
        \item[]{\textbf{Output}:~Merged \ac{LUT} configuration \textit{m}}
        \item[]{\hrulefill}
        \item[]{Number of input pins for configuration \textit{c}: \textit{$c.input\_pin\_size()$}}
        \item[]{Access \textit{i'th} bit of number \textit{n} (MSB at position: 0): $n[i]$}
        \item[]{\hrulefill}
        \While{$s.size() > 1$}
            \State{\textit{$pin := v.take\_first()$}}
            \State{Configuration Vector $\mu_1$, $\mu_2$}
            \State{$\mu_0 := s.take\_first()$}
            \State{$\mu_1 := s.take\_first()$}
            \State{Merged configuration: $merged$}
            \State{$\mu_0\_counter := 0$}
            \State{$\mu_1\_counter := 0$}
            \State{$l := c.input\_pin\_size() + 1$}
            \For{\textit{$n = 0; n < 2^{l}; ++n$}}
                \If{\textit{n[p] == 0}}
                    \State{\textit{$merged.push\_back(\mu_0[\mu_0\_counter++])$}}
                \Else
                    \State{\textit{$merged.push\_back(\mu_1[\mu_1\_counter++])$}}
                \EndIf
            \EndFor
            \
        \EndWhile
        \State{\textit{return $m = s.take\_first()$}}
    \end{algorithmic}
    \label{alg:lut_merge}
\end{algorithm}

\subsection{Trojan implementation}
\label{trojans:section:block_ciphers:trojan}

In order to weaken the provided design we altered the Boolean function for each pattern to return the identity function (input \ac{SBOX} = output \ac{SBOX}), which removes the non-linearity property of PRESENT. Here we again utilized the algorithm \Algorithm{alg:present_pattern_generation} by replacing the standard PRESENT \ac{SBOX} by the identity \ac{SBOX}. During the pattern search phase we now collect all sub-configurations for each LUT-6 element and replace the identified sub-pattern by the equivalent identity \ac{SBOX} configuration based on the detected input pattern. Afterwards the partial equations of the \ac{LUT} needs to be recombined, c.f. \Algorithm{alg:lut_merge} Finally the resulting LUT-6 configuration is written back to the textual netlist representation.

In summary, the presented detection and modification approach allows the modification of any provided \ac{SBOX} implementation in \ac{FPGA} architectures. Complementary to Swierczynski~\etal we demonstrated ways to also handle \ac{SBOX} logic merged with e.g. control logic. In consequence, this allows the detection and weakening of \ac{SBOX} based block ciphers for \ac{FPGA} based designs.

%% file: section/case_study_a51.tex
\section{Case Study: Stream Ciphers}
\label{trojans:section:stream_ciphers}

In this case study, we show how to algorithmically reverse engineer \ac{LFSR}-based stream ciphers (\Section{trojans:section:stream_ciphers:re}) and inject a Trojan into a third-party gate-level netlist based solely on the information inferred from reverse engineering (\Section{trojans:section:stream_ciphers:trojan}).

\subsection{\ac{LFSR}-based Stream Ciphers}
\label{trojans:section:stream_ciphers:re}
Stream ciphers are an important class of widely used encryption algorithms (e.g., in timing-critical voice transmission), since the encryption consists of a simple \textsf{XOR}-ing with a keystream. In practice, various stream ciphers are based on \acp{LFSR} such as E0 in the Bluetooth standard~\cite{bluetooth2016bluetooth}, A5/1 in GSM~\cite{quirke2004security}, or SNOW 3G in UMTS 3G~\cite{orhanou2010snow}. \acp{LFSR} are advantageous since their implementation is lightweight and are mathematically well understood, see~\cite{menezes1996handbook}. In addition, non-linear elements such as a non-linear combination of the output from multiple \acp{LFSR} are used to increase the security of a stream cipher.

\par{\bf A5/1 Stream Cipher.}
The A5/1 algorithm was defined for use in GSM networks and is supposed to guarantee data confidentiality for cellphone calls, however, severe flaws have been detected and various attacks proposed~\cite{biham2000cryptanalysis} ~\cite{biryukov2000real}~\cite{ekdahl2003another}. Besides cryptanalytic attacks, it has been shown that A5/1 can be broken in practice within seconds when using a pre-calculated
table of ciphertext and plaintext pairs (rainbow tables) ~\cite{lu2015time}. Note that our goal is not to demonstrate a cryptanalytic attack by exploiting inherent weaknesses of the A5/1 cipher, but to exemplify how the security of any \ac{LFSR}-based stream cipher can be undermined by exploiting the underlying \ac{LFSR}-based architecture.

\begin{figure}[!htb]
  \centering
  \includegraphics[scale=0.3]{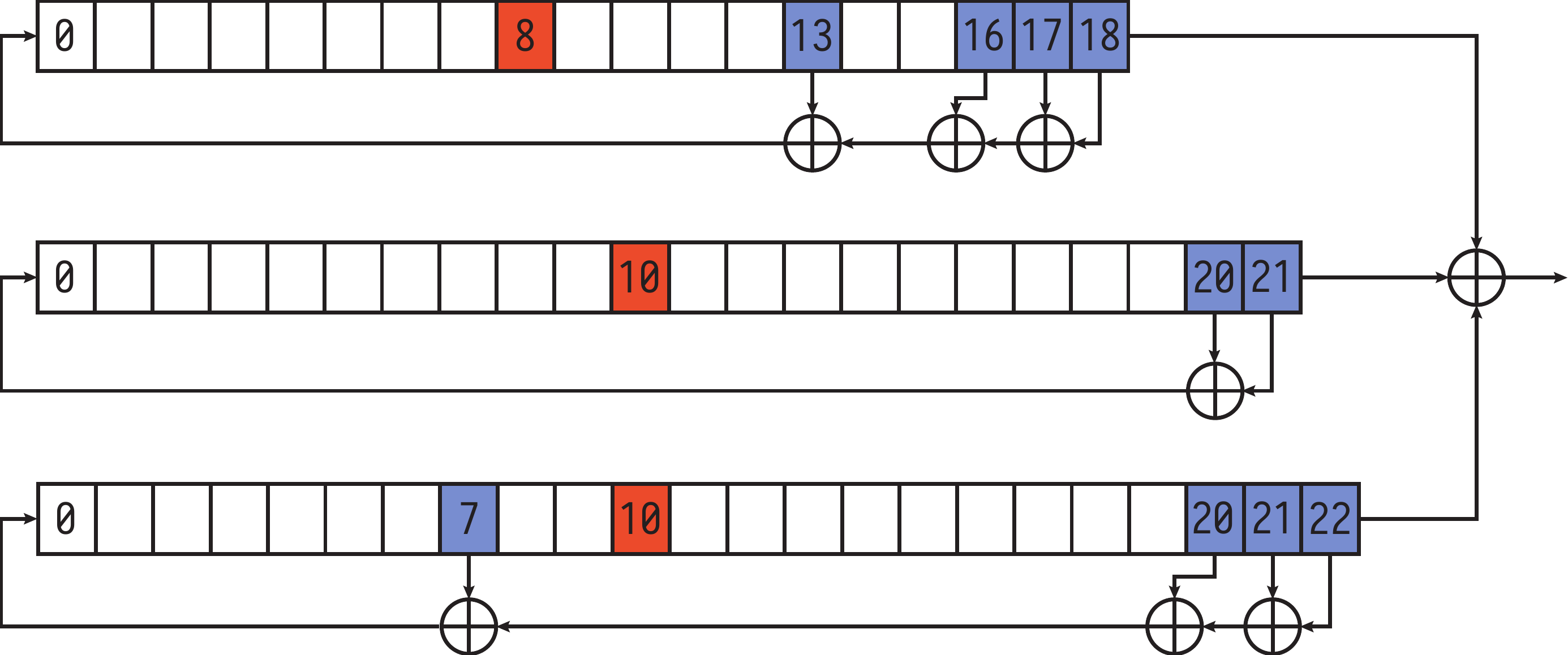}
  \caption{Block diagram of the A5/1 cipher.}
  \label{trojans:figure:a51}
\end{figure}

A5/1's architectural core consists of three \acp{LFSR}, see~\Figure{trojans:figure:a51}.
The outputs of these \acp{LFSR} are \textsf{XOR}-ed and the output in turn is used as bit for the keystream. The \acp{LFSR}, L1, L2 and L3, have lengths of 19, 22 and 21 bits, respectively. To achieve an unpredictable output, they are clocked irregularly. To determine which \ac{LFSR} exactly will be clocked, a so-called majority function is deployed. At positions 8 for L1, 10 for L2 and L3, the bits of the current state are tapped to determine whether 0 or 1 is in the majority. Each \ac{LFSR} that contains the majority bit in aforementioned position will then be clocked.
Before bits for the keystream are generated, a key must be loaded as initial state of the \acp{LFSR}. This ensures the output is unpredictable. Besides a 64-bit key, a public 22-bit frame number is loaded to the \acp{LFSR} by means of feeding the respective bit into all three \acp{LFSR} and subsequently clocking them. This creates an unpredictable initial configuration of the \acp{LFSR}. The majority function is then activated and the \acp{LFSR} are clocked 100 times without any keystream output being produced~\cite{briceno2011pedagogical}.

\par{\bf Reverse Engineering.}
To automatically reverse engineer A5/1 in a gate-level design, we use an extended approach of the \ac{LFSR} detection presented in~\Section{hwre:section:case_wm}. In particular, we search for the specified \ac{LFSR} lengths and the search for the \textsf{XOR} gate that performs the combination of plaintext and keystream. We  have practically verified the correctness of our algorithm on an open-source third-party \ac{IP} core~\cite{Chemeris2008} synthesized with Xilinx ISE 14.7 for an \textsf{xc6slx16}. Note that we integrated the encryption/decryption functionality by addition of the \textsf{XOR} with the plaintext/ciphertext to possess a fully-fledged cryptographic core.

\subsection{Hardware Trojan}
\label{trojans:section:stream_ciphers:trojan}
In order to surreptitiously weaken the cipher, we show how to inject a hardware Trojan into the design's gate-level netlist. The goal of our Trojan is to leak the cryptographic key over the available communication channel.

\par{\bf Trigger.}
As a result of our previous analysis, we know the exact position of the \acp{LFSR}, thus we easily can attach them to our Trojan circuitry. Note that we additionally wire-tap a control signal  which indicate whether the A5/1 core is ready to encrypt or decrypt user data. Once this signal is set to high it triggers our Trojan and we load a copy of the current 64-bit state of the \acp{LFSR} to our Trojan circuitry.

\par
\begin{figure}[!htb]
  \centering
  \includegraphics[scale=0.5]{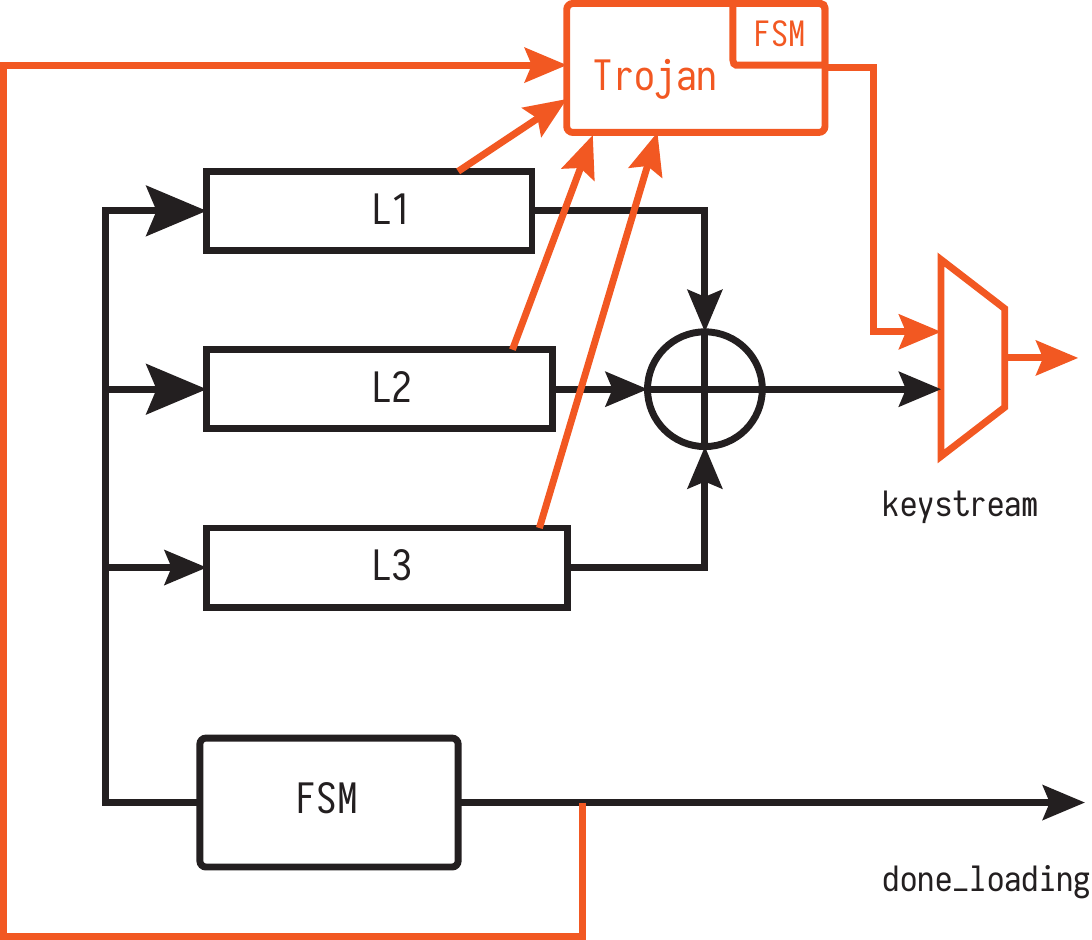}
  \caption{High-level design of the A5/1 Trojan.}
  \label{trojan:figure:a51trojan}
\end{figure}

{\bf Payload.}
The high-level idea of our payload is that the first 64-bit of the ciphertext is the employed 64-bit state of the \ac{LFSR}. To this end, we added a multiplexer after the final \textsf{XOR} gate, see~\Figure{trojan:figure:a51trojan} and a counter to our Trojan. For the first 64 ciphertext bits, we output the stored keystream and thus a man-in-the-middle can decrypt any user data.

One may argue that a user will notice that the encryption failed, as it cannot be decrypted properly. While this is true, we point to the fact that a transmission is organized in bursts of 114 bits. All 4.615 ms ($\sim 217$ times a second) a burst is sent \cite{quirke2004security}, so that the single unreadable burst will likely be unnoticed.

In summary, automated reverse engineering of \ac{LFSR}-based stream ciphers and subsequent hardware Trojan injection is not as challenging as one might think. Due the general nature of our detection algorithm, our approach can be used on basically any \ac{LFSR}-based stream cipher.

%% file: section/conclusion.tex

\section{Conclusion}
\label{trojans:section:conclusion}

Various works have highlighted the threats in modern Integrated Circuits production chain. Diverse threats such as \ac{IP} infringement and injection of hardware Trojans require (partial) reverse engineering of the design.

In this work, we presented how automated reverse engineering supports arbitrary destructive aspects.
First, we provided a reviewed constrained \ac{IP} watermarking scheme for \acp{FPGA} and improved the security by use of opaque predicates. Additionally, we revealed that proposed hardware opaque predicate implementations are not sufficiently secure against reverse engineering.
Second, we provided automated reverse engineering strategies to detect cryptographic implementations and simultaneously inject malicious circuitry in third-party gate-level netlists. Our injected hardware Trojan exploits available communication channels, thus a man-in-the-middle is able to decrypt any communication.

Since our attacks are performed automatically, we believe that our work raises the awareness of the real-world attacker's capabilities targeting cryptographic designs and watermarking schemes.

%% file: main.bbl
\begin{thebibliography}{10}
\providecommand{\url}[1]{#1}
\csname url@samestyle\endcsname
\providecommand{\newblock}{\relax}
\providecommand{\bibinfo}[2]{#2}
\providecommand{\BIBentrySTDinterwordspacing}{\spaceskip=0pt\relax}
\providecommand{\BIBentryALTinterwordstretchfactor}{4}
\providecommand{\BIBentryALTinterwordspacing}{\spaceskip=\fontdimen2\font plus
\BIBentryALTinterwordstretchfactor\fontdimen3\font minus
  \fontdimen4\font\relax}
\providecommand{\BIBforeignlanguage}[2]{{%
\expandafter\ifx\csname l@#1\endcsname\relax
\typeout{** WARNING: IEEEtran.bst: No hyphenation pattern has been}%
\typeout{** loaded for the language `#1'. Using the pattern for}%
\typeout{** the default language instead.}%
\else
\language=\csname l@#1\endcsname
\fi
#2}}
\providecommand{\BIBdecl}{\relax}
\BIBdecl

\bibitem{nist:suite:b}
\BIBentryALTinterwordspacing
NIST, ``{Suite B Cryptography},'' 2001. [Online]. Available:
  \url{https://www.nsa.gov/ia/programs/suiteb\_cryptography/}
\BIBentrySTDinterwordspacing

\bibitem{drimer-thesis}
\BIBentryALTinterwordspacing
S.~Drimer, ``{Volatile {FPGA} design security -- a survey (v0.96)},'' 2008.
  [Online]. Available:
  \url{http://www.cl.cam.ac.uk/~sd410/papers/fpga_security.pdf}
\BIBentrySTDinterwordspacing

\bibitem{cosade:2015:wild}
A.~Wild, A.~Moradi, and T.~G{\"{u}}neysu, ``{Evaluating the Duplication of
  Dual-Rail Precharge Logics on FPGAs},'' in \emph{{COSADE}}, 2015, pp. 81--94.

\bibitem{Sasdrich2014}
P.~Sasdrich and T.~G{\"u}neysu, ``{Efficient elliptic-curve cryptography using
  Curve25519 on reconfigurable devices},'' in \emph{{ARC}}.\hskip 1em plus
  0.5em minus 0.4em\relax Springer International Publishing, 2014, pp. 25--36.

\bibitem{7426148}
F.~Schellenberg, M.~Finkeldey, B.~Richter, M.~Schäpers, N.~Gerhardt,
  M.~Hofmann, and C.~Paar, ``{On the Complexity Reduction of Laser Fault
  Injection Campaigns Using OBIC Measurements},'' in \emph{2015 Workshop on
  Fault Diagnosis and Tolerance in Cryptography (FDTC)}, Sept 2015, pp. 14--27.

\bibitem{5513117}
G.~T. Becker, M.~Kasper, A.~Moradi, and C.~Paar, ``{Side-channel based
  watermarks for integrated circuits},'' in \emph{{HOST}}, June 2010, pp.
  30--35.

\bibitem{karri_hstrust_2012}
R.~Karri, J.~Rajendran, and K.~Rosenfeld, ``{Trojan Taxonomy},'' in
  \emph{{Introduction to Hardware Security and Trust}}, M.~Tehranipoor and
  C.~Wang, Eds.\hskip 1em plus 0.5em minus 0.4em\relax {Springer-Verlag}, 2012.

\bibitem{SurveyTrojan}
M.~Tehranipoor and F.~Koushanfar, ``{A Survey of Hardware Trojan Taxonomy and
  Detection},'' \emph{Design Test of Computers, IEEE}, vol.~27, no.~1, pp.
  10--25, 2010.

\bibitem{5406669}
------, ``{A Survey of Hardware Trojan Taxonomy and Detection},'' \emph{IEEE
  Design Test of Computers}, vol.~27, no.~1, pp. 10--25, Jan 2010.

\bibitem{King:2008:DIM:1387709.1387714}
{S. T. King et al.}, ``{Designing and Implementing Malicious Hardware},'' in
  \emph{{LEET}}, ser. LEET'08.\hskip 1em plus 0.5em minus 0.4em\relax USENIX
  Association, 2008, pp. 5:1--5:8.

\bibitem{ieee:2014:guin}
{U. Guin \etal}, ``{Counterfeit Integrated Circuits: A Rising Threat in the
  Global Semiconductor Supply Chain},'' \emph{Proceedings of the IEEE}, vol.
  102, no.~8, pp. 1207--1228, 2014.

\bibitem{7082768}
J.~Zhang and G.~Qu, ``{A survey on security and trust of FPGA-based systems},''
  in \emph{2014 International Conference on Field-Programmable Technology
  (FPT)}, Dec 2014, pp. 147--152.

\bibitem{book:security_trends_fpga:chapter2}
{E. Wanderley \etal}, \emph{{Security FPGA Analysis}}.\hskip 1em plus 0.5em
  minus 0.4em\relax Springer, 2011, pp. 7--46.

\bibitem{Swierczynski:CAIDI:2015}
P.~Swierczynski, M.~Fyrbiak, P.~Koppe, and C.~Paar, ``{FPGA Trojans Through
  Detecting and Weakening of Cryptographic Primitives},'' \emph{IEEE
  Transactions on Computer-Aided Design of Integrated Circuits and Systems},
  vol.~34, no.~8, pp. 1236--1249, Aug 2015.

\bibitem{bifi}
P.~Swierczynski, G.~T. Becker, A.~Moradi, and C.~Paar, ``{Bitstream Fault
  Injections (BiFI) - Automated Fault Attacks against SRAM-based FPGAs},''
  \emph{IEEE Transactions on Computers}, vol.~PP, no.~99, pp. 1--1, 2017.

\bibitem{moradi:2011:ccs}
{A. Moradi \etal}, ``{On the Vulnerability of FPGA Bitstream Encryption against
  Power Analysis Attacks: Extracting Keys from Xilinx Virtex-II FPGAs},'' in
  \emph{{CCS}}, 2011, pp. 111--124.

\bibitem{moradi:2013:fpga}
------, ``{Side-channel Attacks on the Bitstream Encryption Mechanism of Altera
  Stratix II: Facilitating Black-box Analysis Using Software
  Reverse-engineering},'' in \emph{{FPGA}}, 2013, pp. 91--100.

\bibitem{dt:1999:chisholm}
{G. H. Chisholm \etal}, ``{Understanding Integrated Circuits},'' \emph{{IEEE}
  Design {\&} Test of Computers}, vol.~16, no.~2, pp. 26--37, 1999.

\bibitem{shi:2010:iscas}
{Y. Shi \etal}, ``{A highly efficient method for extracting FSMs from flattened
  gate-level netlist},'' in \emph{{ISCAS}}, 2010, pp. 2610--2613.

\bibitem{li:2013:host}
{W. Li \etal}, ``{WordRev: Finding word-level structures in a sea of bit-level
  gates},'' in \emph{{HOST}}, 2013, pp. 67--74.

\bibitem{gascon:2014:fmcad}
{A. Gasc{\'{o}}n et al.}, ``{Template-based circuit understanding},'' in
  \emph{{FMCAD}}, 2014, pp. 83--90.

\bibitem{hansen:1999:dt}
{M. C. Hansen \etal}, ``{Unveiling the {ISCAS-85} Benchmarks: {A} Case Study in
  Reverse Engineering},'' \emph{{IEEE} Design {\&} Test of Computers}, vol.~16,
  no.~3, pp. 72--80, 1999.

\bibitem{DSB-Report}
``{Report of the Defense Science Board Task Force on High Performance Microchip
  Supply},'' 2005.

\bibitem{fanci}
A.~Waksman, M.~Suozzo, and S.~Sethumadhavan, ``{{FANCI:} identification of
  stealthy malicious logic using boolean functional analysis},'' in \emph{{ACM
  CCS}}, 2013, pp. 697--708.

\bibitem{veritrust}
{J. Zhang et al.}, ``{VeriTrust: Verification for Hardware Trust},''
  \emph{{{IEEE} Trans. on {CAD} of Integrated Circuits and Systems}}, vol.~34,
  no.~7, pp. 1148--1161, 2015.

\bibitem{detrust}
J.~Zhang, F.~Yuan, and Q.~Xu, ``{DeTrust: Defeating Hardware Trust Verification
  with Stealthy Implicitly-Triggered Hardware Trojans},'' in \emph{{ACM CCS}},
  2014, pp. 153--166.

\bibitem{becker_ches_2013}
{G. T. Becker et al.}, ``{Stealthy Dopant-Level Hardware Trojans},'' in
  \emph{{CHES} 2013}, 2013, pp. 197--214.

\bibitem{oakland:2016:yang}
{K. Yanet al.}, ``{{A2:} Analog Malicious Hardware},'' in \emph{{IEEE}
  Symposium on Security and Privacy, {SP} 2016, San Jose, CA, USA, May 22-26,
  2016}, 2016, pp. 18--37.

\bibitem{ghandali:2016:ches}
{S Ghandali et al.}, ``{A Design Methodology for Stealthy Parametric Trojans
  and Its Application to Bug Attacks},'' in \emph{{CHES}}, 2016, pp. 625--647.

\bibitem{DingWZZ13}
{Z. Ding et al.}, ``{Deriving an NCD file from an FPGA bitstream: Methodology,
  architecture and evaluation},'' \emph{Microprocessors and Microsystems -
  Embedded Hardware Design}, vol.~37, no.~3, pp. 299--312, 2013.

\bibitem{tcadics:2001:kahng}
A.~B. Kahng, J.~Lach, W.~H. Mangione-Smith, S.~Mantik, I.~L. Markov,
  M.~Potkonjak, P.~Tucker, H.~Wang, and G.~Wolfe, ``{Constraint-based
  watermarking techniques for design IP protection},'' \emph{IEEE Transactions
  on Computer-Aided Design of Integrated Circuits and Systems}, vol.~20,
  no.~10, pp. 1236--1252, Oct 2001.

\bibitem{ftp:2008:schmid}
M.~Schmid, D.~Ziener, and J.~Teich, ``{Netlist-level {IP} protection by
  watermarking for LUT-based FPGAs},'' in \emph{{FTP}}, 2008, pp. 209--216.

\bibitem{ecr:2006:ginger}
G.~Myles and C.~Collberg, ``Software watermarking via opaque predicates:
  Implementation, analysis, and attacks,'' \emph{Electronic Commerce Research},
  vol.~6, no.~2, pp. 155--171, 2006.

\bibitem{sergeichik:2014:jicms}
V.~Sergeichik and A.~Ivaniuk, ``{Implementation of Opaque Predicates for FPGA
  Designs Hardware Obfuscation},'' \emph{Journal of Information, Control and
  Management Systems}, vol.~12, no.~2, 01/2014.

\bibitem{Bogdanov:2007:PUB:1421964.1422007}
A.~Bogdanov, L.~R. Knudsen, G.~Leander, C.~Paar, A.~Poschmann, M.~J. Robshaw,
  Y.~Seurin, and C.~Vikkelsoe, ``{PRESENT: An Ultra-Lightweight Block
  Cipher},'' in \emph{{Proceedings of the 9th International Workshop on
  Cryptographic Hardware and Embedded Systems}}, ser. CHES '07.\hskip 1em plus
  0.5em minus 0.4em\relax Berlin, Heidelberg: Springer-Verlag, 2007, pp.
  450--466.

\bibitem{present-opencore}
\url{https://opencores.org/project,present}.

\bibitem{jcen:2016:swierczynski}
{P. Swierczynski \etal}, ``{Interdiction in Practice---Hardware Trojan Against
  a High-Security USB Flash Drive},'' \emph{{Journal of Cryptographic
  Engineering}}, pp. 1--13, 2016.

\bibitem{bluetooth2016bluetooth}
S.~Bluetooth, ``{Bluetooth specification version 5},'' \emph{Available HTTP:
  http://www. bluetooth. com}, 2016.

\bibitem{quirke2004security}
J.~Quirke, ``{Security in the {GSM} system},'' \emph{{A}us{M}obile, May}, pp.
  1--26, 2004.

\bibitem{orhanou2010snow}
G.~Orhanou, S.~El~Hajji, and Y.~Bentaleb, ``{SNOW 3G stream cipher operation
  and complexity study},'' \emph{Contemporary Engineering Sciences-Hikari Ltd},
  vol.~3, no.~3, pp. 97--111, 2010.

\bibitem{menezes1996handbook}
A.~Menezes, P.~van Oorschot, and S.~Vanstone, ``{Handbook of Applied
  Cryptography. CRC, 1996},'' ISBN 0-8493-8523-7, Tech. Rep.

\bibitem{biham2000cryptanalysis}
E.~Biham and O.~Dunkelman, ``{Cryptanalysis of the A5/1 GSM stream cipher},''
  in \emph{{International Conference on Cryptology in India}}.\hskip 1em plus
  0.5em minus 0.4em\relax Springer, 2000, pp. 43--51.

\bibitem{biryukov2000real}
A.~Biryukov, A.~Shamir, and D.~Wagner, ``{Real Time Cryptanalysis of A5/1 on a
  PC},'' in \emph{{International Workshop on Fast Software Encryption}}.\hskip
  1em plus 0.5em minus 0.4em\relax Springer, 2000, pp. 1--18.

\bibitem{ekdahl2003another}
P.~Ekdahl and T.~Johansson, ``{Another attack on A5/1},'' \emph{IEEE
  transactions on information theory}, vol.~49, no.~1, pp. 284--289, 2003.

\bibitem{lu2015time}
J.~Lu, Z.~Li, and M.~Henricksen, ``{Time--memory trade-off attack on the {GSM}
  {A}5/1 stream cipher using commodity {GPGPU}},'' in \emph{{International
  Conference on Applied Cryptography and Network Security}}.\hskip 1em plus
  0.5em minus 0.4em\relax Springer, 2015, pp. 350--369.

\bibitem{briceno2011pedagogical}
M.~Briceno, I.~Goldberg, and D.~Wagner, ``A pedagogical implementation of a5/1
  (1999),'' \emph{Dostupn{\'e} na: http://www. scard. org/gsm/a51. html}, 2011.

\bibitem{Chemeris2008}
\url{https://github.com/chemeris/airprobe-gprs/blob/master/A5.1/Verilog/a5.v}.

\end{thebibliography}
